\documentstyle[12pt]{article}
\begin{document}
\setcounter{page}{0}
\begin{flushright}
{CERN--TH/95--139}
\end{flushright}
\vskip0.5cm
\begin{center}

{\large\bf CP-Violating Effect of QCD Vacuum \\
\vskip0.2cm
in Quark Fragmentation}

\vskip1cm

A. Efremov$^{a,}$\footnote{E-mail: efremov@thsun1.jinr.dubna.su.}
and D. Kharzeev$^{b,c,}$\footnote{E-mail: kharzeev@vxcern.cern.ch}\\

\vskip0.8cm
{\it $^a)$ Laboratory of Theoretical Physics\\
Joint Institute for Nuclear Research\\
Dubna, 141980 Russia}
\medskip

{\it $ ^b)$ Theory Division\\
CERN\\
Geneva, Switzerland}
\medskip

{\it $^c)$ Fakult\"at f\"ur Physik\\
Universit\"at Bielefeld\\
33615 Bielefeld, Germany}
\end{center}
\vskip1cm
\begin{abstract}
We demonstrate that the non-trivial CP-violating structure of QCD vacuum
can lead to an observable effect in $e^+e^-\to q\bar q\to 2$-jet
annihilation. We find that the sign of the jet handedness correlation
can be opposite
to that predicted by factorization of $q~\bar q$ fragmentation and
CP-conjugation of the two jets. A simple model estimation of the handedness
correlation magnitude is given.
\end{abstract}

\vspace {-0.10in}

\vskip3.5cm

CERN--TH/95--139

May 1995

\newpage

{\bf 1.} The problem of CP violation in strong interactions is of fundamental
interest. Even though CP-violating effects are not expected in perturbative
QCD, the non-trivial structure of QCD vacuum may lead to the breakdown
of the invariance with respect to CP conjugation \cite{CP}.
A particular illustration of this option is the appearance of the
so-called $\Theta$-term in the effective Lagrangian of strong interactions:
\begin{equation}
\L_{\Theta} = {\Theta \over{16\pi^2}} {\rm tr} (G_{\mu\nu}\tilde{G}^{\mu\nu}),
                                                                  \label{CP}
\end{equation}
which is intimately related to the possibility of tunnel transitions
between the vacuum states with different topological indices \cite{tun}.
This tunnelling phenomenon is invoked, for example, in the now generally
accepted solution of the $U(1)$ problem in QCD \cite{u1}.
It is easy to see that the term (\ref{CP}) explicitly breaks down CP
invariance. However CP-violation in strong interactions was never observed.
Does it mean that the true theory of strong interactions is CP-invariant?
\vskip0.3cm

The answer to this question is not necessarily positive, since
in the usual observables the non-perturbative CP-violating effects
would appear only as tiny corrections, which are difficult to isolate.
A possibility of CP-violation can be connected with solutions of
classical equations of
motion in gluodynamics with a nonzero field in the ground state~\cite{dual}. It
was  suggested that such self-dual fields could provide color confinement and
linearly rising Regge trajectories~ \cite{efim}.  In fact, the existence of
non-vanishing vacuum gluon fields assumed in the  QCD Sum Rule
method~\cite{SVZ} would definitely induce the CP-violating effects due
to a chromo-magnetic component of considerable strength:

\begin{equation}
\langle \alpha_s B^2 \rangle = {1 \over 4} \langle \alpha_s G^2 \rangle
\simeq 0.01\ {\rm GeV}^4,
                                                               \label{SVZ}
\end{equation}
where at a later stage we have used the ``canonical" gluon condensate
value~\cite {SVZ}. However the terms induced by the vacuum gluon fields
generally
enter the expressions for physical quantities as higher-twist power
corrections to the leading perturbative QCD results; these corrections are
proportional to the square of the gluon field strength tensor
$\langle \alpha_s G^2 \rangle$. It is extremely important therefore to
find an observable
in which the non-perturbative effects would give a {\it leading} contribution.
\vskip0.3cm

Such an observable was recently proposed in the work of ref. \cite{meribel}:
it is
the jet handedness correlation in the $e^+e^-\to Z^0\to$ 2-jet process.
There is a preliminary experimental indication \cite{meribel} that this
correlation is of the sign opposite to that predicted by CP-conjugation
of $q$ and $\bar q$ jets and by their independent fragmentation.
\vskip0.3cm

The aim of this Letter is to demonstrate that the observed sign is natural
if the fragmentation occurs in the background of non-perturbative
chromo-magnetic vacuum field.  We shall present a simple model estimation of
the handedness correlation.
\vskip0.3cm

{\bf 2.} Let us consider the process of $e^+e^-$ annihilation into two jets:
\begin{equation}
e^+e^- \rightarrow q + \bar{q} \rightarrow jet_{q} + jet_{\bar{q}}\ .
                                                         \label{jet}
\end{equation}
Since the $\bar{q}q$ pair is produced by the (spin 1) virtual photon or
in the $Z^0$ decay, the spins of the quark and antiquark will be aligned
in the same direction (opposite helicities). Consider now the pairs of positive
and negative pions formed in the fragmentation of the jet. Let us call the pair
``left" (``right") if the projection of the vector product of the three-momenta
of positive and negative pions on the thrust vector $\vec{t}$ in the
direction of the jet (i.e. the azimuthal angle $\phi$ of the pair with
respect to the jet direction)
\begin{equation}
{(\vec{k}_+ \times \vec{k}_-) \vec{t} \over{|\vec{k}_+| |\vec{k}_-|}}=
\sin\phi
\label{phi}
\end{equation}
is negative (positive). CP conjugation transforms a ``left" (``right")
pair produced in the quark-induced jet into the ``left" (``right") pair in
the antiquark-induced jet, and vice versa. The ``handedness"~\cite{hand} of
the jet can be defined as
\begin{equation}
H = {N_R - N_L \over{N_L + N_R}},                                  \label{Hs}
\end{equation}
where $N_L$ ($N_R$) is the number of ``left-handed" (``right-handed") pairs of
charged pions formed in the fragmentation of the jet. This quantity is
shown~\cite{hand} to be proportional to the parent quark polarization $P_q$.

Let us now define also the ``handedness correlation" parameter as
\begin{equation}
C = {N_{LR} + N_{RL} - N_{RR} - N_{LL} \over{N_{LR} + N_{RL} + N_{RR} +
N_{LL}}},                                                       \label{corr}
\end{equation}
where $N_{LR}$, for example, stands for the number of events in which
the pair in the quark-induced jet was ``left" but the pair in the
antiquark-induced jet was ``right".
\vskip0.3cm

Naively, one would expect that the fragmentation of quark and antiquark jets
is CP-symmetric, as it is in perturbative QCD. This means that a pair of
some handedness in one jet has to correspond mainly to the pair of the same
handedness in the opposite jet since the variable (\ref{phi}) is CP--even,
i.e. the correlation (\ref{corr}) has to be negative. This can be well
illustrated in a simple classical model where the handedness arises from the
turning of secondary quark and antiquark produced in breaking of string in
a longitudinal chromo-magnetic field from chromo-magnetic dipole moments of
initial $q\mbox{ and }\bar q$ \cite{ryskin}. Since the spin directions of
initial $q$ and $\bar q$ are the same the directions of the chromo-magnetic
fields are opposite. This results in the same handedness in both jets
and in the negative
correlation (\ref{corr}). There exists, however, an experimental indication
\cite{meribel} that it is definitely positive. This should
mean that the CP--conjugation of the two jets is broken.
\vskip0.3cm

As we shall now discuss, an entirely non-perturbative component of
the fragmentation -- a vacuum gluon field, first considered in the context
of jet physics in ref. \cite{VZ} -- could induce a local CP-parity violation
and hence could lead to a positive value of $C$. Indeed, in the spirit of the
model~\cite{ryskin} the quark and antiquark ``shaken-off" in the fragmentation
process will be deviated by the background vacuum chromo-magnetic field
(\ref{SVZ}) in the same direction, both in the quark- and antiquark-induced
jets. The unit vector $\vec n\propto (\vec{k}_+ \times \vec{k}_-)$ will thus
have a length different from zero and will be aligned in the same direction in
both jets. This will induce an increase in the number of events with
{\it different }``handedness" in opposite jets, $N_{LR}$ and $N_{RL}$, leading
to the positive value of the ``handedness correlation" (\ref{corr}).
\vskip0.3cm

{\bf 3.} Let us now present these qualitative arguments
in a somewhat more formal way. The two-particle fragmentation function of a
polarized quark fragmenting into a $(+-)$-pair in the background
chromo-magnetic field $\vec{B}^a$ can be written down in the Lab system as

\begin{equation}
D^B_q = w_q [1 + \alpha_q (\vec{s}\vec{n}) + \beta_q (\vec{B}^a\vec{n})],
                                                                \label{frag}
\end{equation}
where $\vec{s}$ is the spin of the quark, $w, \alpha$ and $\beta$ depend on
the longitudinal and transverse (with respect to the thrust axis) momenta
$k_{\pm}^{L,T}$ of the particles and on the field strength $B^2$. Under the
charge conjugation, $\vec{n}$ and $\vec{B}$ change sign, but $\vec{s}$ does
not. Due to the charge conservation of fragmentation $D_q^B=D_{\bar q}^{-B}$
and
\begin{equation}
\alpha_{\bar{q}} = - \alpha_{q},  \ \beta_{\bar{q}} = \beta_q.
\end{equation}
Averaging over different events with presumably random orientation of $\vec{B}$
and over azimuthal angle of $\vec{n}$, one obtains for handedness and
handedness correlation in $e^+e^-\to$ 2-jet annihilation the following
expressions (assuming naturally that $\langle \vec{B} \rangle =0$):
\begin{equation}
H^{e^+e^-} = {\sum_q\sigma_qw_q \alpha_q P_q
\over \sum_q\sigma_q w_q}\  \mbox{ and }\
C^{e^+e^-} = {\Sigma_q \sigma_q w_q^2 (-\alpha_q^2 c_{q\bar{q}} +
\beta_q^2 \langle B^2 \rangle)
\over{ \Sigma_q \sigma_q w_q^2}},                            \label{correl}
\end{equation}
where $P_q$ is the longitudinal quark polarization, $c_{q\bar{q}}=+1$ is the
$q\bar{q}$ helicity correlation defined through the difference of the number of
$q\bar{q}$ pairs with opposite helicities and with the same one and $\sigma_q$
is the production cross section of the flavour $q$. Formula (\ref{correl})
shows explicitly that the presence of a chromo-magnetic component of the gluon
condensate of QCD vacuum can result in the positive sign of the handedness
correlation (if the third term in (\ref{frag}) prevails)
and hence to CP-asymmetry in the fragmentation of the two jets.
Therefore, even though the parameter (\ref{corr}) is CP-even, its sign can
indicate the presence of CP-asymmetry of the fragmentation process in QCD
vacuum.
\vskip0.3cm

{\bf 4.} We shall now try to make some quantitative estimates.
Assuming that the transverse size of the vacuum fluctuation $l$ and the
transverse momentum $k_T$ of the ``shaken-off" quark and antiquark satisfy
the condition
\begin{equation}
k_T l >> 1,                                                  \label{cond}
\end{equation}
one can estimate the magnitude of the effect in the classical approximation
\cite{VZ,ryskin}. Using (\ref{frag}), it is easy to see that the
``handedness" (\ref{Hs}) due to chromo-magnetic field $B$ is related to the
angle $\chi$ of the deviation of quark (antiquark) from the back-to-back axis
by the following relation, which is valid when the angle $\chi$ is small and
one can use a linear (in $\sin\chi$) approximation to the antisymmetric part
of the distribution of the pair in the azimuthal angle $\phi$:

\begin{equation}
H_B =\langle{\rm sign}(\sin\phi)\rangle \simeq
{4 \over \pi} \langle\sin\phi\rangle =
{4\over {\pi}} \langle{\rm sin} 2\chi \rangle\simeq {8 \over\pi}
\langle \chi\rangle,             \label{hi}
\end{equation}
where $\langle ...\rangle$ means averaging over the azimuthal angle. The
``handedness correlation" parameter (\ref{corr}) in the presence of the
field $B$ is given by
\begin{equation}
C_B\simeq -(4/\pi)^2 \langle{\rm sin} \phi \rangle_{q-jet} \langle{\rm sin}
\phi'
\rangle_{\bar{q}-jet}.
\end{equation}
Since we consider the case of interaction with vacuum
fields,
$$\langle{\rm sin} \phi \rangle_{q-jet} = - \langle{\rm sin} \phi'
\rangle_{\bar{q}-jet},$$
and we obtain
\begin{equation}
C _B\simeq {16\over \pi^2}  \langle{\rm sin} \phi \rangle_{q-jet}^2
\simeq H_B^2.
\label{square}
\end{equation}

Using the classical equation of motion of quark in the external
chromo-magnetic field, we can write down the following expression for the mean
deviation angle:
\begin{equation}
\langle \chi \rangle \simeq {\Delta k \over {k_T}}
\simeq {g B_{||} \tau \over{k_T}}.
                                                             \label{angle}
\end{equation}
Here $\tau$ is the proper "formation time" of the meson
\footnote{Recall that according to the uncertainty principle it is a minimal
time during which a virtual fluctuation with energy deficit $\Delta E$ is
undistinguishable from the initial state.}, i.e. the  time during which the
massless quark feels the chromo-magnetic field {\it before} it transforms to a
color-neutral meson. In the jet center-of-mass system, it is given
by~\cite{ftime}

\begin{equation}
\tau \simeq {1\over 2\epsilon_{\rm c.m.}}={1 \over{2k_T}\cosh(y-Y)},
\label{tau}
\end{equation}
where $\epsilon_{\rm c.m.}$ is the meson energy in the jet c.m. whereas $y$ and
$Y={1\over2}\ln((E_{jet}-P_{jet})/(E_{jet}+P_{jet}))$ are the rapidities of the
meson and of the center-of-mass of the jet in Lab system, respectively.

Turning to the correlation it is necessary to take into account that the
fragmentation of $q$ and $\bar q$ takes place in different space-time points
$x$ and $x'$, and therefore the product of fields in different points will
appear. These fields
are not necessarily located in the same vacuum field domain. Averaging over
all possible configurations of the domains restores the Lorentz invariance,
making the product of the fields dependent on the interval $(x-x')^2$. The
averaging also results in a suppression of the bi-local product
when the
interval becomes larger than some
correlation length $l$ (a ``domain size") \cite{efim}.

Recalling that for the longitudinal component of the chromo-magnetic field one
has
\begin{equation}
\langle g^2 B_{||}(x) B_{||}(x') \rangle = {1\over 3} \langle
g^2\vec B(x) \vec B(x') \rangle =
{\pi \over 3} \langle \alpha_s G(x) G(x') \rangle,
\end{equation}
we obtain for the handedness correlation from (\ref{square}), (\ref{hi}) the
following expression\footnote{We do not write down explicitly the Schwinger
phase factor in the gluon field correlator required by gauge invariance.}:
\begin{equation}
C \simeq {64 \over {3\pi}} \langle \alpha_s G(x) G(x') \rangle
{\tau\tau' \over k_Tk_T'}\ ,
\end{equation}
where the brackets $\langle...\rangle$ denote averaging over all
possible configurations of
the domains.
\vskip0.3cm

The vacuum domain size is poorly known at present; to get an idea about
the value of $l$ we can turn to the results of lattice calculations
\cite{Di}, which show that for large {\it Euclidian} space-time
intervals $R^2=(t'-t)^2+(\vec{r}-\vec{r}')^2$ the correlators $\langle
G(x)G(x') \rangle$
fall off exponentially,
\begin{equation}
\langle G(x) G(x') \rangle \simeq \langle G^2(0) \rangle
\exp{\left(-{\sqrt{R^2}\over l}\right)}\ ,
\end{equation}
with $l\simeq 0.2$ fm. We will assume that this behavior
holds for space-like intervals $(x'-x)^2<0$ in Minkowski space-time since
the change
$t\to it$ results in the substitution

$$ R^2 \to -(t'-t)^2 + (\vec{r}'-\vec{r})^2 = -(x'-x)^2\simeq 4tt',$$

\noindent
where  at the last stage we have taken into account that the points $x$ and
$x'$, over
which the average is performed, are located in opposite fragmentation regions
and quarks are moving approximately with the velocity of light.
We thus get for space-like
intervals

\begin{equation}
C \simeq {64 \over {3\pi}} \langle G^2(0) \rangle
\exp{\left(-2{\sqrt{tt'}\over l}\right)}
{\tau\tau' \over k_Tk_T'}\ .            \label{result}
\end{equation}

Using for the gluon condensate the value (\ref{SVZ}) and assuming
for the $k_T$ an average value $\langle k_T\rangle \simeq 0.3\ GeV/c$
we find from (\ref{result}) for the maximal value of the handedness correlation
parameter at $\sqrt{tt'}\simeq l \simeq 0.2\ fm$ the value of the order of

\begin{equation}
C \simeq {64 \over {3\pi}} \langle G^2(0) \rangle e^{-2}
\left({l\over \langle k_T\rangle\gamma}\right)^2
\simeq+0.5\%,                                         \label{estim}
\end{equation}
where $\gamma =E_{jet}/M_{jet}\simeq 9$ is the Lorentz-factor for
transformation from the Lab system to the jet center of mass system.
\vskip0.3cm

This value seems to underestimate the preliminary experimental observation
\cite{meribel}. However we have to stress once more that, first, we use a
simple
classical model which is hardly good for small $k_T$, second, the
correlation length $l$ is poorly known (some estimations~\cite{efim}
give for it the value of up to $0.6\ fm$),  and third, the
observation~\cite{meribel} is very preliminary and selection of events made
there is very severe.  We would like to stress nevertheless in conclusion,
that the positive value of $C$ corresponds to the CP-violating effect and this
very non-trivial phenomenon deserves further experimental and
theoretical
investigation.

In particular, if it is really an effect of a vacuum field it should be
accompanied also by an asymmetry corresponding to a vacuum {\it
chromo-electric} field approximately of the same strength. It is not
difficult to show that there has to be an asymmetry with respect to
the difference of velocities of particles in pairs. Indeed, the covariant form
of the third term in~(\ref{frag}) is $G_{\mu\nu}k_+^\mu k_-^\nu$. In the
product of the fragmentation functions for $q$ and $\bar q$ averaged over
the vacuum field $G$ this results in a term
\begin{equation}
\langle GG' \rangle \left[(k_+{k'}_+)(k_-{k'}_-) -
(k_+{k'}_-)(k_-{k'}_+)\right],
\label{covar}
\end{equation}
because the only covariant and P-invariant form of the vacuum matrix element
is
$$
\langle G_{\mu\nu}(x){G}_{\mu'\nu'}(x') \rangle ={1\over12} \langle GG' \rangle
\left(g_{\mu\mu'}g_{\nu\nu'}-g_{\mu\nu'}g_{\nu\mu'}\right)\ .
$$
The expression (\ref{covar}) can be rewritten as
\begin{equation}
\langle GG' \rangle {\epsilon}_+{\epsilon'}_+{\epsilon}_-{\epsilon'}_-
\left[(\vec v_+\times\vec v_-)({\vec v'}_+\times{\vec v'}_-) -
(\vec v_+ - \vec v_-)({\vec v'}_+ - {\vec v'}_-)\right]\ ,
\label{velcor}
\end{equation}
where $\vec v = \vec k/\epsilon$ is the velocity of a particle in a pair.
While the first term in (\ref{velcor}) is responsible for "chromo-magnetic"
correlation, the longitudinal part of which has been considered above, the
second one has to produce a "chromo-electric" correlation: the difference of
velocity for particles in a pair in one jet should prefer to be directed in
the opposite hemisphere to the difference in the opposite jet. It would be
interesting
to look for this effect experimentally.

\vskip0.3cm

The authors are grateful to J. Ellis, P. Minkowski, H. Satz, O. Teryaev, L.G.
Tkatchev, V.I. Zakharov and to the referee of the first version of the
paper for helpful discussions. A. E. acknowledges the hospitality of CERN
Theory Division during the initial stage of this work. The work of A.E. was
supported in part by the International Science Foundation under Grant  FE300,
by the INTAS Grant 93-1180 and by the Russian Foundation for Fundamental
Research under Grant 93-02-3811. D. K. acknowledges financial support of the
German Research Ministry (BMFT) under contract 06 BI 721.


\end{document}